\documentclass[twocolumn,showpacs,showkeys,superscriptaddress]{revtex4}
\usepackage{graphicx}
\usepackage{bm}
\usepackage{color}
\usepackage{amsmath}
\usepackage{natbib}
\usepackage{latexsym}
\usepackage{mathrsfs}

\begin{document}

\title{ Resonant Energization of Relativistic particles by an Intense Electromagnetic Wave}
\author{Swadesh M. Mahajan}
\email{mahajan@mail.utexas.edu}
\affiliation{Institute for Fusion Studies, The University of Texas at Austin, Texas 78712, USA.}
\date{\today}

\begin{abstract}

The phenomenon of resonant energization of a relativistic quantum particle, moving in unison with an intense ElectroMagnetic Wave, is demonstrated in a semiclassical calculation. The wave nature of the quantum particle is of essence because the resonant process originates in wave-wave interaction-between the classical EM wave, and the quantum wave associated with the particle. When the energy /momentum of the quantum wave satisfy the resonance condition (the effective phase speeds of the two waves are equal), the particle, drawing energy directly from the intense EM field, could acquire extremely high energies  Such a direct resonant energy transfer from intense Electromagnetic  waves will constitute a hitherto unexploited mechanism that could power the most energetic of cosmic rays. Some predictions of the theory will, hopefully, be tested in the laboratory Laser experiments.
\end{abstract}

\pacs{03.65Pm; 03.65Sq; 52.27.Ny; 52.35.Mw; 52.38.-r}

\keywords {Wave-Wave Interaction, Quantum Particle Waves, Intense Electromagnetic Waves, Lasers, Resonant Energization, super energy particles}

\maketitle

Let us begin with a deep but relatively trivial realization that the "group velocity" of the "quantum wave" associated with a relativistic particle (with the energy-momentum relation, $E^{2}= c^{2} P^2+m^{2}c^4$)
\begin{equation} \label{phasespeed}
v_{p}= \frac {\partial E}{ \partial P} = c^2 \frac{P}{E}= c \frac{P}{(P^2+m^{2}c^{2})^{1/2}}
\end{equation}
tends to approach $c$ as the particle momentum becomes much greater than its rest mass. Under appropriate conditions, then, such a particle could resonantly exchange energy/momentum with a propagating electromagnetic (EM) wave. This direct ( and persistent) transfer of energy/momentum from a very high intensity EM wave to the resonant particles could drive them to extremely high energies.  

It is suggested that, in a cosmic setting where relativistic particles find themselves in the vicinity of an intense source of EM energy (emanating from some cataclysmic event), most energetic of cosmic rays could be generated by this mechanism.

What is remarkable is that the principal qualitative, and some quantitative aspects of this possible spectacular phenomenon can be derived, illustrated and demonstrated in a simple straightforward semi-classical calculation (describing  the particle quantum mechanically while the intense EM wave is treated classically). The quantum treatment of the particle dynamics is crucial since the entire phenomenon results from the interaction of two waves - the EM and the quantum wave associated with the particle.  

Needless to say there are any number of high energy laser laboratories that are equipped to test the  fundamental predictions of Resonant energization. I will propose an appropriate experiment in the concluding part.

I will get straight to this demonstration using the theoretical framework developed in a recent paper \cite{MA-16}
that attempts to build ``A Statistical Model for Relativistic Quantum Fluids interacting with an Intense Electromagnetic Wave" based on the solutions of the eigenvalue problem embodied in the relativistic quantum equations - the Klein-Gordon (KG) and the Dirac equation- in the presence of a given arbitrary amplitude EM wave. It is important  to  point out that the literature is quite full of exact and approximate solutions similar to those described in MA-16. In fact, the attempts to solve the KG-EM and Dirac-EM systems began almost ninety years ago \cite{Gordon,Volkov}. The Volkov vacuum solution of the Dirac-EM system become the reference for a variety of later papers in which the solution was extended to a plasma \cite{Redmond,FelbMar,NN76,Beck77,VolkovMen,Boca,RaiElie,Varro1,HAK16,KingHu,Seipt,MS19}; aside from Volkov-type results, other solutions have also been derived for these systems \cite{BG05,BG14}. There is considerable overlap between the KG/Dirac solutions  presented in \cite{RaiElie} and MA-16 (the scope of MA-16 is much larger than just the solutions of the wave equations -it is  the "construction" of a relativistic quantum fluid based on these solutions) that both could be considered  basic references for this note.
 
Despite the fact the Resonant solutions (the subject matter of this paper) were implicitly present in \cite{RaiElie}, MA-16 (and perhaps other works), no one, including this author, noticed/emphasized their special and amazing nature; these eigenstates, with highly boosted energy-momentum, are  a unique expression of a resonant wave-wave interaction- between the quantum KG (Dirac) wave and an intense classical EM wave propagating together. This paper is dedicated to delineation and explanation of this remarkable resonance and predicting some of the consequences of these  boosted energy states. 

Before presenting the structure of the wave-wave resonance energization, it must be pointed put the acceleration of particles (including the so called Direct laser Acceleration (DLA)) in intense laser fields is a very highly/deeply investigated subject (theory and simulation) \cite{LaserAccDLA}- Though highly impressive, these representative papers focus  on solving the Lorentz equation of a classical particle in the EM field. The phenomenon that we are about to dwell in this paper, however, is dictated by the quantum (wave-like) behavior of the "particle", and is not  readily accessible to purely classical treatments mentioned above. So the vast literature on laser induced acceleration is not directly relevant to the content of this paper. 

Although the solutions of the  Dirac equation are considerably more complicated (and richer) than the KG solutions, the essence of the phenomenon of resonant energization, barring some inessential details,  can be fully captured in a KG analysis. I will, therefore, present only the much simpler KG calculation in some detail and  simply state the corresponding results for the Dirac system. For analytic simplicity, the EM wave is assumed to be circularly polarized. The contra-variant components of electromagnetic four potential $A^\mu$ of a CPEM, propagating in the $z$-direction, are (The Minkowski signature tensor $\eta^{\mu\nu}=\mbox{diag}[1,-1,-1,-1])$
\begin{equation} \label{cpem1}
A^0=0=A^z,\\ A^x=A \cos(\omega t-k z), A^y=-A \sin(\omega t-k z),   
\end{equation}
where $A$ is its constant amplitude, and the four wave vector $k^{\mu}=[\omega,0,0,k]$ in the lab frame. Notice that $k^{\mu} A_{\mu}=0 =\bf {k}\cdot \bf {A}$, and the Lorentz invariant  $A^\mu A_\mu= A^2$ has no space time dependence; it is the latter property that makes  CPEMs most amenable to analytical calculations. The CPEM electric ${\bf E}= (\omega A(\hat e_x \sin(\omega t-k z) +\hat e_y \cos(\omega t-k z)))$ and magnetic $({\bf B}= k A(-\hat e_x \cos(\omega t-k z) +\hat e_y \sin(\omega t-k z)))$ fields imply a {\it constant} Poynting flux (${\bf E\times\bf B}= k \omega A^2 \hat e_z$) along z.

For most details the reader is referred to MA-16. The KG equation for a particle of rest mass $m$ and charge $q$, interacting with the electromagnetic field ($\hbar=1=c$), is given by
\begin{equation}\label{KGMin2}
 \left (p_\mu+qA_\mu\right)  \left(p^\mu+qA^\mu\right)\Psi=\left (i\partial_\mu+qA_\mu\right)  \left(i\partial^\mu+qA^\mu\right)\Psi=m^2\Psi
\end{equation}
where $p^\mu=[p^0=E, \bf p]$  is the energy momentum four vector of the particle. For the CPEM of (\ref{cpem1}), Eq.(\ref{KGMin2}) expands as
\begin{equation}\label{KGExpand2}
 \partial_t^2\Psi-\nabla^2\Psi-2iqA \left[\cos(\omega t-k z)\partial_x\Psi-\sin(\omega t-k z)\partial_y\Psi\right]+\mathcal{M}^2\Psi=0
\end{equation}
where $\mathcal{M}=\sqrt{m^2+q^2A^2}$ is the field-renormalized effective mass, a concept of great significance emphasizing that the EM wave contributes profoundly to the effective inertia of the quasi-quantum particle (QPF) born out of the KG/EM union. Equation (\ref{KGExpand2}), then, can be viewed as the wave equation of the QPF of mass $\mathcal{M}$, still in  interaction with the CPEM through the term proportional to $qA$. 

Out of the four space-time dimensions, the two space dimensions orthogonal to the propagation direction ($z$) are ignorable. We can, then, define the  independent Fourier mode
\begin{equation}\label{fourier1}
 \Psi(x,y,z,t)= e^{iK_\perp(\cos\varphi x+\sin\varphi y)} \psi(t,z)
\end{equation}
that obeys
\begin{equation}\label{psi/zt}
 \partial_t^2\psi- \partial_z^2\psi+2qAK_\perp\cos(\omega t-k z)\psi+ ({K_\perp} ^2+\mathcal{M}^2)\psi=0
\end{equation}
where $K_x=K_\perp\cos\varphi$, $K_y=K_\perp\sin\varphi$, and the phase $\varphi$ has been eliminated in a trivial redefinition of the variable $z$. One notes:
\begin{itemize}
\item The perpendicular momentum $K_\perp$ is a good quantum number. Thus $A$ and $K_\perp$ are simply specified numbers. 
\item The QPF dynamics (\ref{psi/zt}), just like the EM dynamics, is controlled by the hyperbolic wave operator ($ \partial_t^2-\nabla^2$) - the analogous mathematical structure is precisely what can make the two waves resonate; it provides the necessary condition for the existence of the resonant wave-wave interaction, the principal  mechanism for energy transfer. Evidently, such a phenomenon will be hard to recover for classical particles.
\item the dependence of the interaction term in (\ref{psi/zt}) on $z$ and $t$ implies that neither the energy ($E$) nor the z-momentum ($K_z$) of the QPF is a constant of the motion. We will, later, define  the respective expectation values, $<E>$ and $<K_z>$. 
\item The explicit interaction term in (\ref{psi/zt}) vanishes for $K_\perp=0$. In this limit, both $E$ and $K_z$ become good quantum numbers, and (\ref{psi/zt}) allows the exact solution   
 \begin{equation}\label{Fid1}
 \Psi(z,t)= \Psi (0)e^{-i E t +iK_{z} }
\end{equation}
provided the EM- field modified relativistic dispersion relation
 \begin{equation}\label{Fid2}
E^2=K_z^2+\mathcal{M}^2= K_z^2+{m}^2+q^2A^2
\end{equation}
is satisfied. This exact result, though rather simple, is quite noteworthy: When the motion of the quantum particle is limited to the direction of the EM propagation, the particle "energy" increases by $q^2A^2$ over its field-free value; no other discernible change can be seen. This special solution will be used as a Fiducial /Reference Solution in the rest of this paper. 

\end{itemize}

Of all the classes of solutions accessible to (\ref{psi/zt}), I will concentrate here only on those that have a direct bearing on the resonant energy exchange between the two waves - the quantum particle wave, and the classical EM wave. Though the phenomenon of Resonant Energization has some similarities with another well known resonant process (called Landau damping  in plasma physics \cite {Landau}),  it is quite different in its nature, context, as well as in its consequences. Landau damping is generally a linear process, caused when a self consistently generated wave (often electrostatic) damps on a "conventional" classical particle (imparting energy to the particle in the process); both the wave and the particle are strictly classical. In the Resonant Energization process, reported in this paper, the energy exchange takes place between two waves: the  quantum KG wave and the classical EM wave. As pointed out earlier, the mathematical roots for the resonant energy exchange lie in the hyperbolic wave operator ($ \partial_t^2-\nabla^2$) that controls  the dynamics of both the KG and the Maxwell equations. 

The best route to finding the resonant solution is to stipulate that $\psi$ depends on t and z through the combination $\zeta= (\omega t-k z)$, the phase factor of the EM wave. Equation (\ref{psi/zt}), then, converts to ($\psi=\psi(\zeta)$)
\begin{equation}\label{psi/zeta}
(\omega^2-k^2)\frac{d^2\psi}{d\zeta^2}+[\mu+\lambda \cos\zeta]\psi=0
\end{equation}
where $\mu=K_\perp^2+m^2+q^2A^2$ and $\lambda=2qAK_\perp$ ( $\lambda/\mu <1$). Equation (\ref{psi/zeta}) is a simple Mathieu equation and can be rigorously analyzed. However, a physically motivated ``approximate" analysis is much more accessible and transparent.

The most striking characteristic of (\ref{psi/zeta}) is that the highest derivative term is multiplied by $\omega^2-k^2$; this coefficient is strictly zero for an EM wave propagating in vacuum. In that case, the exact system allows only the trivial solution $\psi=0$.  However, if the EM dispersion relation were of the form $\omega^2-k^2=w^2$ with  $w\ll\omega$ (as for a tenuous plasma), Equation (\ref{psi/zeta}) is a perfectly well-defined but a singular differential equation- singular in the sense that it will support a solution for which  $d^2\psi/d\zeta^2$ is sufficiently large to balance the other terms. Since large $d\psi/d\zeta$ implies larger energy and z-momentum, this will constitute the Resonant solution that we set out to seek. I will now give much needed details to complete the picture. 

Let us first study the $K_\perp=0$ (the fiducial case) limit of (\ref{psi/zeta}) [$\lambda\rightarrow 0, \mu\rightarrow m^2+q^2A^2$]. From the  (exact) positive energy solution,  
\begin{equation}\label{ResBasic}
\psi=\psi(0) e^{-i\sqrt{{\frac{m^2+q^2A^2}{\omega^2-k^2}}}\\ (\omega t-k z)}, 
\end{equation}
one readily calculates the particle energy, and z-momentum 
 \begin{equation}\label{ResEn}
<E>= \frac {i}{\psi}{\frac {\partial\psi}{\partial t}} = (m^2+q^2A^2)^{1/2}  \frac {\omega}{\sqrt{\omega^2-k^2}},
\end{equation}
\begin{equation}\label{ResMo}
<K_z>= -\frac {i}{\psi} {\frac {\partial\psi}{\partial z} } = (m^2+q^2A^2)^{1/2} \frac {k}{\sqrt{\omega^2-k^2}}.
\end{equation}
We have put the symbols $< >$ around $E$ and $K_z$ to emphasize that we are dealing with expectation values despite the fact in the limit of $K_\perp=0$, both energy and momentum are exactly conserved.

Evidently, both $<E>$ and $<K_z>$ become very large as $\omega\rightarrow k$; the explicit expression of the energy/momentum enhancement (boost)factor 
\begin{equation}\label{ResEnhance}
\frac {\omega}{\sqrt{\omega^2-k^2}} \gg 1
\end{equation}
is one of the more important technical result of this paper. 
 
It may appear somewhat odd  that the two exact solutions (\ref{Fid1})-(\ref{Fid2}), and (\ref{ResBasic})-(\ref{ResMo}) of the same ($K_\perp=0$) equation seem to look so different. I will now show that the solution (\ref{ResBasic})-(\ref{ResMo}) is entirely consistent with (\ref{Fid1})-(\ref{Fid2}) but has deeper content. From  (\ref{ResEn}) and (\ref{ResMo}), one may readily derive
  \begin{equation}\label{Fid22}
<E^2>-<K_z>^2= (m^2+q^2A^2) [ \frac {\omega^2}{\omega^2-k^2}- \frac {k^2}{\omega^2-k^2}]=m^2+q^2A^2,
\end{equation}
exactly the same energy-momentum relationship as in (\ref{Fid2}). This, of course, had to be true - it is simply the generic relativistic energy momentum relation in the presence of a CPEM. But the remarkable attribute of the solution (\ref{ResBasic})-(\ref{ResMo}) is that we have been able to calculate explicit, independent expressions for both $<E>$ and $<K_z>$ where as  (\ref{Fid1})-(\ref{Fid2}) is limited only to the energy-momentum relationship (\ref{Fid2}). It is these explicit formulas that demonstrate the fundamental phenomenon of resonantly boosted energy/momentum of the QPF; the calculation yields an explicit expressions for the boost factor $\mathcal{E}_b$ to boot!

Further insight into the deeper content of the solution (\ref{ResBasic})-(\ref{ResMo}) is obtained when one finds that the ratio of the QPF energy to its z-momentum, 
 \begin{equation}\label{ratio}
\frac{<E>}{<K_z>}=  \frac{\omega}{k}
\end{equation}
is exactly equal to the ratio of the energy of the EM wave to its z-momentum; one could hardly imagine a more compelling signature of a resonant phenomenon! The equality (\ref{ratio}) leads to the same exponents in (\ref{Fid1}) and (\ref{ResBasic}) establishing formal equivalence.

I will now briefly deal with the full equation (\ref{psi/zeta}) to derive somewhat detailed structure of the Resonant Solution that reduces to (\ref{ResBasic})-(\ref{ResMo}) in the limit $K_\perp=0=\lambda$. Let us recall that  the resonant energy (measured in units of the rest mass $m$) $E (K_\perp=0)=\sqrt{(m^2+(qA)^2}\omega/ \sqrt{\omega^2-k^2}$ is much greater than unity which means that the $\zeta$ variation contained in (\ref{ResBasic}) is much stronger than the $\zeta$ variation of the interaction term ($\cos\zeta$) in (\ref{psi/zeta}). This suggests a WKB analysis (see MA-16) but we will follow a simple perturbation approach to show that the essentials of the Resonant solution are already contained in (\ref{ResBasic})-(\ref{ResMo}). Let (see the definition of $\mu$ following  (\ref{psi/zeta}))
\begin{equation}\label{ResFull}
\psi=\psi_s(\zeta) e^{-i{{\frac{\sqrt\mu}{\sqrt{\omega^2-k^2}}}} \\ \zeta}\equiv \psi_s(\zeta) e^{-iS \zeta}
\end{equation}
where $\psi_s(\zeta)$ represents the slow variation [compared to the eikonal, i.e, $(d\psi_{s}/{d\zeta})/(\psi_{s}) \ll S=\sqrt {\mu/ (\omega^2-k^2)}$] of the wave function. Substituting (\ref{ResFull}) into (\ref{psi/zeta}) and neglecting ${d^2\psi_{s}}/{d\zeta^2}$, we derive 
\begin{equation}
\frac{1}{\psi_{s}}\frac{d\psi_{s}}{d\zeta}=\frac{i\lambda}{2S (\omega^2-k^2)}\cos\zeta
\end{equation}
that integrates to
\begin{equation}
\psi_{s}=e^{ i{\frac{2qA}{2[\mu(\omega^2-k^2)]^{1/2}}\\ \sin\zeta}}=e^{i\alpha \sin\zeta}
\end{equation}
from which it is trivial to verify that $d\psi_{s}/{d\zeta}=|\alpha|\ll S$ even when $K_\perp$ is not small. Thus the approximate but complete solution 
\begin{equation}
\psi_{s}=e^{-iS\zeta+i\alpha \sin\zeta}= \sum_n J_n(\alpha) \psi_n, \\  \psi_n=e^{-iE_n t+iK_{zn}z}
\end{equation}
appears to be a sum of n terms. The nth component occurs with a weight $J_n(\alpha)$ and has characteristic energy  and z-momentum, 
\begin{equation}
E_n= \sqrt {\mu} \frac {\omega}{\sqrt{\omega^2-k^2}}-n\omega,  \\  K_{zn}=\sqrt {\mu} \frac {k}{\sqrt{\omega^2-k^2}}-nk,  
\end{equation}
Since $\sqrt {\mu/(\omega^2-k^2)}\gg1$, $E_n$ ($K_{zn}$) can differ substantially from $<E>(<K_z>)$ only for very large $n$ where the Bessel functions tend to be rather small for moderate values of $\alpha$. Extensive numerical solutions of the Mathieu equation (\ref{psi/zeta}) reveal that apart from some minor modulations (evidently reflecting the higher n components discussed above) the basic structure of the Resonant Solution is well represented by the set (\ref{ResBasic})-(\ref{ResMo}) with the eikonal S going from  $\sqrt {\mu/(\omega^2-k^2)}$ towards $\sqrt {(\mu+\lambda)/(\omega^2-k^2)}$ as $K_\perp$ becomes larger, and $\lambda$ approaches $\mu$

In order to advance further, we must find a reasonable evaluation/estimate for $w^2=(\omega^2-k^2)$, the quantity that controls the energy amplification factor of the Resonant solution. Remembering that in this analysis, the arbitrary amplitude EM wave is not only externally specified but it has also been treated as a monochromatic wave with $\omega$ and $k$ as parameters determining the wave phase. In vacuum propagation $w=0$, and the preceding analysis has to be redone. But let us assume that the CPEM wave is passing through a charged gas (plasma) of weakly interacting QPF so that the induced current can be calculated simply by summing the individual contributions of these  KG particles. We can, then, find the self-consistent solution of  the Maxwell equations to determine the dispersion relation obeyed by the EM wave ; this dispersion relation, would relate $\omega$ and $k $.  For the relativistic KG plasma (with rest frame density $n_R$ and temperature T), such a dispersion relation was calculated, for instance, in MA-16 (Eq. 46),   
\begin{equation}\label{DispRel}
\omega^{2} - k^2 = \frac{{\omega_p}^2}{\sqrt{(1+q^2A^2/m^2)} \Gamma_{th}}
\end{equation}
where $\omega_{p}=\sqrt{4\pi q^{2}n_R/m}$ is the invariant (rest frame) plasma frequency, and $\Gamma_{th} ( = K_2(g)/K_1(g)  >1$ for a relativistic Maxwellian plasma) signifies the thermal enhancement of the particle inertia. The argument  $g=\sqrt{(m^2+q^2A^2)}/T$ is the inverse of the normalized temperature- but normalized to the field enhanced inertia of the QPF and not to the bare rest mass. In the rest of this paper,  $\Gamma_{th} =1$.

The dispersion relation (\ref{DispRel}) gives us the necessary input to derive explicit expressions for  normalized energy( $\Gamma$) and normalized z-momentum ($K$) (with $K_\perp\neq 0$) fully in terms of the "specified " quantities (plasma density,  Intensity and the frequency of the CPEM field)
\begin{equation}\label{ResEnF}
\Gamma=\frac{<E>}{m} = [(1+\frac{q^2A^2+K_\perp^2}{m^2})^{1/2}][ 1+\frac{q^2A^2}{m^2})^{1/4} \frac{\omega}{\omega_p}] 
\end{equation}
\begin{equation}\label{ResMoF}
K=\frac{<K_z>}{m}= [(1+\frac{q^2A^2+K_\perp^2}{m^2})^{1/2}][1+\frac{q^2A^2}{m^2})^{1/4} \frac{k}{\omega_p}] 
\end{equation}
The  EM field boosts up the normalized energy and z-momentum through two distinct processes: 1) by contributing, non-resonantly, field energy to perpendicular kinematic energy (the first square bracket), and 2) through the resonant enhancement epitomized in the factor $[1+q^2/A^2)/m^2]^{1/4} {\omega}/{\omega_p}$; the latter can become arbitrary large (even for moderate values of $qA/m$) since in many cases of interest, $\omega\gg \omega_p$. There is an equally spectacular gain (through the combined effects of the resonant and non-resonant energization) for strong EM fields ($qA\gg m$), the normalized energy eventually going up as $(qA/m)^{3/2}$. 

It is, perhaps, worthwhile to reiterate that the only those QPF waves that satisfy the resonant condition $<E>/<K_z> =  \omega/k$ (propagating  in unison with the EM wave along z direction) will, preferentially, gain  large amounts of  z-momentum from the EM wave. And it is this gain that translates into a commensurate gain in QPF energy while maintaining the relativistic energy momentum relationship $<E>^2=<K_z>^2+ K_\perp^2+\mathcal{M}^2$ where $\mathcal{M}^2 = (m^2+q^2A^2)$ reflects the field-modified effective mass. 

Let us now put in perspective the straightforward (almost effortless) demonstration of the rather spectacular phenomenon of direct resonant transfer of energy-momentum from an intense EM wave to a relativistic quantum particle.  Of course, the reverse process of the EM wave gaining energy from the relativistic particles ( analogue of Landau Growth) is, theoretically, just as possible under appropriate conditions. An interesting example of the inverse mechanism at work can be seen in \cite{ML18} where an EM wave is shown to feed off the free energy in a sheared velocity field. 

I will, now, spell out  the basic assumptions, develop an overall scenario, and state the most important potential consequences of this work:
 \begin{itemize}

\item Although the calculations presented here were derived for arbitrary parameters ( $m$, $qA$, $\omega$, $k$,   $<K_\perp>$--), they are, subliminally, geared towards highly relativistic particles and intense EM fields, i.e,  $<E>/m \simeq <K_z>/m\gg1$, and $qA/m\gg1$. For typical EM waves of interest (Intense Optical Lasers, for example), on the other hand, the energy `` per quantum" (normalized to , say, the electron rest mass) is rather small: $\omega/m \simeq k/m\ll1)$; clearly the typical particle or field energy  is much much larger than $\omega$. These ratios and the and the enhancement factor $(1+(qA/m)^2)^{1/4}\omega/\omega_p\gg1$ are needed to estimate the ``extent" of resonant energization.

\item Since no energy inventory is presented, it is assumed that the EM wave acts as an intense constant energy pump. Such a situation could, readily, be created in very high energy laser experiments. One can also imagine that this calculation will pertain to super strong energization of cosmic ray particles that happen to be in the vicinity of a cataclysmic event (colliding blackholes, for example) that is accompanied by prodigious emissions of electromagnetic energy. Since the basic mechanism is the resonance between a class of relativistic particle-waves and a wave that propagates essentially with the speed of light, it is hardly a stretch to conclude that intense gravitational waves ( produced again in some cataclysmic event) could just as well substitute EM waves. 

\item The quantum wave investigated in this paper is the KG wave that pertains to a spin less particle. The analysis is easily extended to a half spin Dirac particle ( calculations are a lot more involved). However, the essentials of the Resonant energization process are more or less the same ( there is very little difference in the KG and Dirac cases in the aforementioned regime when  $<E>/\omega\gg1$. More detailed solution of the KG, a complete solution of the Dirac equation, and the equivalent calculations for the gravitational waves  will be the subject matter of  forthcoming papers. Mostly analytical calculations will be supplemented by numerical solutions. 

\item One must, at least, comment on the probability of finding a particle that will be subjected to such Resonant Energization. I believe that the answer to this question requires considerable thought but the resonant condition ( $<E>/<K_z>=  \omega/k$) does provide a clue - the resonant particles will be found only a small but non-zero sliver of the  $K_\perp - K_z$ phase space.

\item This paper is limited to a conceptual formulation and delineation of the Resonant Energization mechanism. However, I will make some simple estimates for the efficacy of this process when  the normalized energies are high. For $qA/m\gg1, qA\gg K_\perp$, $\Gamma\approx (qA/m)^{3/2}(\omega/\omega_p)$. Let us examine the consequences of this formula in two different contexts (where conditions for Resonant energization  might prevail):

1) High intensity laser plasma experiments:  Consider a wave of amplitude $qA/m\approx {10^2-10^4}$ propagating in a barely under-dense plasma with $\omega/\omega_{p} \geq1$. The Resonant solutions then correspond to a total normalized energy $\Gamma= E/m \sim (qA/m)^{3/2}\sim 10^3-10^6$ of which the resonant boost factor $\Gamma_{res}\approx (qA/m)^{1/2}\approx 10-10^2$). Two straightforward experiments be proposed to look for the signatures of the proposed mechanism: a) qualitative-Searching for particles with "unexpectedly" high energies, b) quantitative- by testing the predicted scaling of the resonant energy with the wave amplitude $\Gamma\approx(qA/m)^{3/2}$. 

2) In a cosmic setup, the plasma densities will be much smaller and $\omega/\omega\gg1$. One could, then, expect to  generate a proton in the PEV energy range, $\Gamma_{proton}\approx 10^6$) by the combination $qA/M_{p} =10$ and $\omega/\omega_{p}=10^5$ \cite {HEnCos}. With this mechanism, it is eminently possible to catapult protons to much higher energies since EM  fields associated with cataclysmic events could be of a prodigious magnitude.
 
\end {itemize}
To the best of my knowledge, this is the first exposition of the existence and characteristics of super energetic particle states created by the resonant transfer of energy-momentum from intense EM (or gravitational waves) waves to relativistic particles-waves propagating in unison. This profound mechanism, then, should be seriously investigated as one of the processes that may be powering the most energetic cosmic rays.

\begin{acknowledgements}
I am very grateful to Felipe Asenjo  for continual support and discussions. I also thank Manasvi Lingam for a critical reading of the manuscript. This work was partially supported by the  US-DOE grants DE-FG02-04ER54742.    

\end{acknowledgements}

{}


\begin{thebibliography}{1}

\bibitem{MA-16} S.M. Mahajan, F. A. Asenjo, Phys. Plasmas {\bf 23}, 056301(2016)

\bibitem{Gordon} W.Gordon,  Z. Phys. {\bf 40}, 117-133 (1927)

\bibitem{Volkov} D. M. Volkov, Z. Phys. {\bf 94}, 250 (1935)

\bibitem{Redmond} Peter J. Redmond,  Journal of Mathematical Physics {\bf 6}, 1163 (1965)- This paper solves the  relativistic KG-EM and Dirac-EM
systems in the presence of a static magnetic field along the direction of propagation. 

\bibitem{FelbMar} Franklin S. Felber, and John H. Marburger, Journal of Mathematical Physics, 16, 2089 (1975)

\bibitem{NN76} N. B. Narozhnyi and A. I. Nikishov, Theoretical and Mathematical Physics, {\bf 26}, 9 (1976)

\bibitem{Beck77} W. Becker, Physica A, {\bf 87}, 601 (1977)
 
\bibitem{VolkovMen} J. T. Mendonca and A. Serbeto, Phys. Rev. E {\bf 83}, 026406 (2011),
J. T. Mendonca, Phys. Plasmas {\bf 18}, 062101 (2011)

\bibitem{Boca} M. Boca, J. Phys. A, {\bf 44}, 445303 (2011)

\bibitem{RaiElie}Erez Raicher, and Shalom Eliezer, Phys. Rev A, {\bf 88}, 022113 (2013)

\bibitem{Varro1} Sandor Varro, Laser Phys.Lett {\bf 10}, 09531(2013),\\
Sandor Varro, Laser Phys.Lett {\bf 11}, 016001(2014), \\Sandor Varro, Nuclear Instruments and Methods in Physics Research {\bf A 740} (2014) 280?283

\bibitem{HAK16} T. Heinzl, A. Ilderton and B. King, Phys. Rev. D, {\bf 94}, 065039 (2016)

\bibitem{KingHu} B. King and H. Hu, Phys. Rev. D, {\bf 94}, 125010 (2016)

\bibitem{Seipt} D. Seipt, Proceedings of the Summer School "Quantum Field Theory at the Limits: From Strong Fields to Heavy Quarks" , arXiv:1701.03692 (2017)

\bibitem{MS19} J. T. Mendonca and A. Serbeto, Europhys. Lett., {\bf 125}, 65002 (2019)

\bibitem{BG05} V. Bagrov and D. Gitman, Annalen der Physik, {\bf 517}, 467 (2005)

\bibitem{BG14} V. G. Bagrov and D. Gitman, \emph{The Dirac Equation and its Solutions}, De Gruyter, Berlin (2014)

\bibitem {LaserAccDLA} 
J. Pang, Y. K. Ho, X. Q. Yuan, N. Cao, Q. Kong, P. X. Wang, L. Shao,
E. H. Esarey and A. M. Sessler, Phys.Rev.E {\bf 66}, 066501( ??2002??)
 
P. X. Wang, Y. K. Ho,a) X. Q. Yuan, Q. Kong, N. Cao, L. Shao, A. M. Sessler, E. Esarey, E. Moshkovich, Y. Nishida, N. Yugami, and H. Ito, J. X. Wang and S. Scheid

A. P. L. Robinson, A. V. Arefiev, and D. Neely, PRL {\bf111}, 065002 (2013)

A. V. Arefiev, V. N. Khudik, P. L. Robinson, G. Shvets, L. Willingale, and M. Schollmeier, Phys. Plasmas {\bf 23}, 056704 (2016);

Tianhong Wang, Vladimir Khudik, Alexey Arefiev, and Gennady Shvets, arXiv:1812.052099v1[Physics.plasma-ph], 13 Dec 2018
and references therein

\bibitem {Landau} Realizing that there are far too many references to Landau, we give here a representative few: 

LD Landau, On the vibrations of the electronic plasma. Journal of Physics (USSR) {\bf10(1)}:25-34 (1946),
Lifshitz EM, Pitaevskii LP, Physical Kinetics, Pergamon, Oxford (1981). 

In relativistic plasmas: 

Volokitin, A. S., Krasnoselskikh, V. V. and Machabeli, G. Z. {\bf11}, 531 (1985),
Mahajan, S., Machabeli, G., Osmanov, Z.  and Chkheidze, Nat. Sci. Rep. {\bf 3}, 1262 (2013).

\bibitem {ML18} S. Mahajan and M. Lingam, Phys. Plasmas, {\bf 25}, 072112 (2018)
\bibitem{HEnCos} Abramowski, A., Aharonian, F., Benkhali, F. A., et al. (H.E.S.S. Collaboration) 2016, Natur, 531, 476,   Z Osmanov, S Mahajan, and G Machabeli, The Astrophysical Journal, 835:164 (4pp), 2017. 

\end{thebibliography}
\end{document}